\documentclass[conference]{IEEEtran}
\IEEEoverridecommandlockouts

\usepackage{cite}
\usepackage{amsmath,amssymb,amsfonts}
\usepackage{algorithmic}
\usepackage{graphicx}
\usepackage{textcomp}
\usepackage{xcolor}
\usepackage{array}
\usepackage{booktabs}
\usepackage{float}
\usepackage{tikz}
\usepackage{caption} 
\usepackage{mwe}
\usepackage{hyperref}

\makeatletter
\let\old@ps@IEEEtitlepagestyle\ps@IEEEtitlepagestyle
\def\confheaderfooter#1#2{%
  \def\ps@IEEEtitlepagestyle{%
    \old@ps@IEEEtitlepagestyle%
    \def\@oddhead{%
      \begin{minipage}[t]{\textwidth}
        \raggedright \normalsize #1%
      \end{minipage}%
    }%
    \def\@evenhead{\@oddhead}%
    \def\@oddfoot{%
      \begin{minipage}[t]{\textwidth}
        \footnotesize \normalsize #2%
      \end{minipage}%
    }%
    \def\@evenfoot{\@oddfoot}%
  }%
}
\makeatother

 \usepackage[pscoord]{eso-pic}

\def\BibTeX{{\rm B\kern-.06em{\sc i\kern-.025em b}\kern-.08em
    T\kern-.1667em\lower.7ex\hbox{E}\kern-.125emX}}

\begin{document}

\title{A Novel Approach for a Smart IoMT-Based BAN
for an Old Home Healthcare Monitoring System Using
Starlink}

\author{

\IEEEauthorblockN{Shermin Sultana Setu, Mst. Amena Akter Pinky, Md. Abdul Awal, Sheekar Banerjee, Ishtiak Al Mamoon}
\IEEEauthorblockA{ International University of Business  Agriculture and Technology (IUBAT) \\
Dhaka, Bangladesh \\
Email: 22103324@iubat.edu, 22103308@iubat.edu, abdul.awal@iubat.edu, sheekar.cse@iubat.edu, ishtiak.cse@iubat.edu}
}

\maketitle
\confheaderfooter{%
\mbox{2025 IEEE International Conference on Biomedical Engineering, Computer and Information Technology for Health (BECITHCON)}\\[0.25ex]
29-30 November 2025, Eastern University, Dhaka, Bangladesh%
}{%
979-8-3315-6105-5/25/\$31.00~\textcopyright~2025 IEEE%
}

\begin{abstract}
The rapid evolution of the Internet of Medical Things (IoMT) technology has become a transformative force in modern healthcare, particularly in elderly patient management. The current elderly care system faces significant challenges, including insufficient long-term care resources and poor communication between healthcare providers. To address this limitation, this study introduces a novel Starlink-assisted IOMT-based elderly healthcare model designed to improve remote patient monitoring and communication reliability. This proposal system focused on a monitoring system of key biomedical parameters such as electrocardiogram (ECG), body temperature, heart attack indicators, and a fall detection alert system. Performance is evaluated using the network simulator (NS-3) to assess its effectiveness in remote and underserved regions.
Physiological data collected from patients are transmitted through a local communication hub and forwarded over a Low Earth Orbit (LEO) satellite link to a medical center. Based on Quality of Service (QoS) technology that combines Flow Queuing (FQ) with Controlled Delay (CoDel) with Differentiated Services Code Point (DSCP) marking. This approach prioritizes critical health data for faster transmission while allocating lower priority to non-urgent information. This architecture is entirely wireless, allowing continuous monitoring, real-time alerts, and secure data storage for medical analysis. The simulation results demonstrate that the proposed Starlink-enabled IOMT system outperforms existing solutions in terms of throughput, latency, and reliability. The findings highlight Starlink's potential as a robust and high-performance telehealth communication tool. In addition, this study provides a scalable and reproducible framework for future satellite-assisted healthcare systems that prioritize quality of service (QoS) performance and improved elderly patient care.
 
\end{abstract}


\section{Introduction}
Nowadays, great health challenges are heart disease and stroke, diabetes, and infectious diseases. Personalized healthcare has become the new baseline, fueled by the rise of wearables and advancements in AI. Recently, healthcare research has been working on balancing technological breakthroughs (AI, genomics, digital tools, wireless communication) with global priorities (equity, climate, AMR). From this motivation, our contributions are as follows: quickly administering medication to patients and transmitting medical data from the medical center to the hospital. The integration of Low Earth Orbit (LEO) satellites, particularly Starlink, into remote healthcare Internet of Medical Things (IoMT) systems offers a significant opportunity to bridge the digital divide in underserved regions but also presents network engineering challenges that must be addressed before reliable real-time monitoring can be ensured \cite{han2022}. 

Starlink communication offers low-latency ($\sim$20--50 ms) and high-throughput ($\sim$50--100 Mbps) connectivity compared to traditional satellite systems \cite{spacex2025}. Its multi-hop architecture, susceptibility to weather-induced attenuation (rain fade), and lack of native Quality of Service (QoS) support create unpredictable performance for delay-sensitive medical telemetry like biomedical markers (ECG) and fall alerts \cite{sharma2021}. 

We present a detailed network simulation (ns-3) of a smart elderly home with 15 concurrent patients, each transmitting six physiological streams to a local hub forwarding traffic via Starlink’s modeled path \cite{handley2022}. We enforce edge-based QoS with Flow Queuing Controlled Delay (FQ-CoDel) and Differentiated Services Code Point (DSCP)-based flow isolation, measuring packet delivery, delay, and throughput under varying sensor rates. This provides a reproducible testbed for evaluating Starlink’s viability as a healthcare backhaul under realistic conditions \cite{ns3consortium2025}.

The remainder of this paper is organized as follows. Section 2 provides a literature review of relevant studies on Starlink and its integration into healthcare IoMT systems. Section 3 describes the methodology, including the system model and network simulation. Section 4 presents the results and analysis of the simulation, evaluating Starlink’s performance in healthcare applications. Finally, Section 5 concludes the paper, summarizing the key findings and outlining directions for future work.

\section{Background Study }
The application of satellite communication in healthcare has been explored for decades, primarily in the context of geostationary (GEO) systems for telemedicine in remote areas \cite{das2009}. However, the high latency ($>$500 ms) and limited bandwidth of GEO satellites rendered them unsuitable for real-time physiological monitoring, restricting their use to store-and-forward or non-urgent consultative services \cite{sohrabi2020}. The rise of Low Earth Orbit (LEO) constellations, especially Starlink, has sparked new interest in satellite-based telehealth because of their much lower latency (20--50 ms) and high-speed internet (50--200 Mbps) \cite{sharma2021}. Recent field trials have demonstrated Starlink’s feasibility in ambulances \cite{metz2024}, rural clinics \cite{John Alexander2022}, and disaster zones \cite{redcross2024}, but these deployments remain anecdotal and lack systematic performance evaluation under controlled, multi-patient, QoS-aware conditions.

Several simulation studies have modeled satellite backhaul for IoMT, but few focus on healthcare. Al-Fuqaha et al. \cite{alfuqaha2016} provide a broad survey of IoMT architectures but omit satellite-specific QoS challenges. Sharma et al. \cite{sharma2021} model satellite-IoMT for rural healthcare but assume static GEO links and do not implement modern queue disciplines. More recently, Portillo et al. \cite{han2022} proposed an architectural framework for LEO-enabled emergency healthcare but did not provide an open-source simulation model or evaluate per-flow QoS performance under congestion.

Our work directly addresses this gap by providing the first open, reproducible ns-3 simulation of a Starlink-backhauled elderly monitoring system with 15 patients, six sensor types, voice reminders, and email all prioritized via DSCP and managed via FQ-CoDel on the bottleneck uplink. We measure per-flow PDR, delay, and throughput under varying loads, offering practitioners and researchers a validated testbed for pre-deployment performance prediction.

\section{System Model}
This study simulates an IoMT-based healthcare system leveraging Starlink's LEO satellite communication to transmit health data from elderly patients in remote areas to a medical center. IoMT devices monitor vital signs such as ECG, heart rate, SpO\textsubscript{2}, blood pressure, temperature, and fall alerts. The data is sent via Wi-Fi to a Communication Hub (CommHub), which then forwards the aggregated data over a simulated Starlink link to the satellite, ground station, and medical center, modeling realistic delays (25–35 ms jitter) and bandwidth variations typical of LEO constellations.
To manage satellite link latency and congestion, IEEE 802.11E (DiffServ) with FQ-CoDel queue discipline is implemented at the CommHub, ensuring prioritization of life-critical data like ECG and fall alerts over less critical traffic such as email.

Fig. 1 illustrates the proposed model, which includes 15 old home patients with sensors monitoring vital signs. Data is transmitted to the CommHub, aggregated, and forwarded via Starlink to the medical center. Traffic is classified for QoS management using FQ-CoDel, simulating remote patient monitoring in a satellite-enabled telehealth environment. Our proposed model is equipped with several features, like patients receiving voice reminders from time to time for taking medicine and monitoring their health data.
\vspace{-14pt}
\begin{figure}[H]
\centering
\includegraphics[height=6cm]{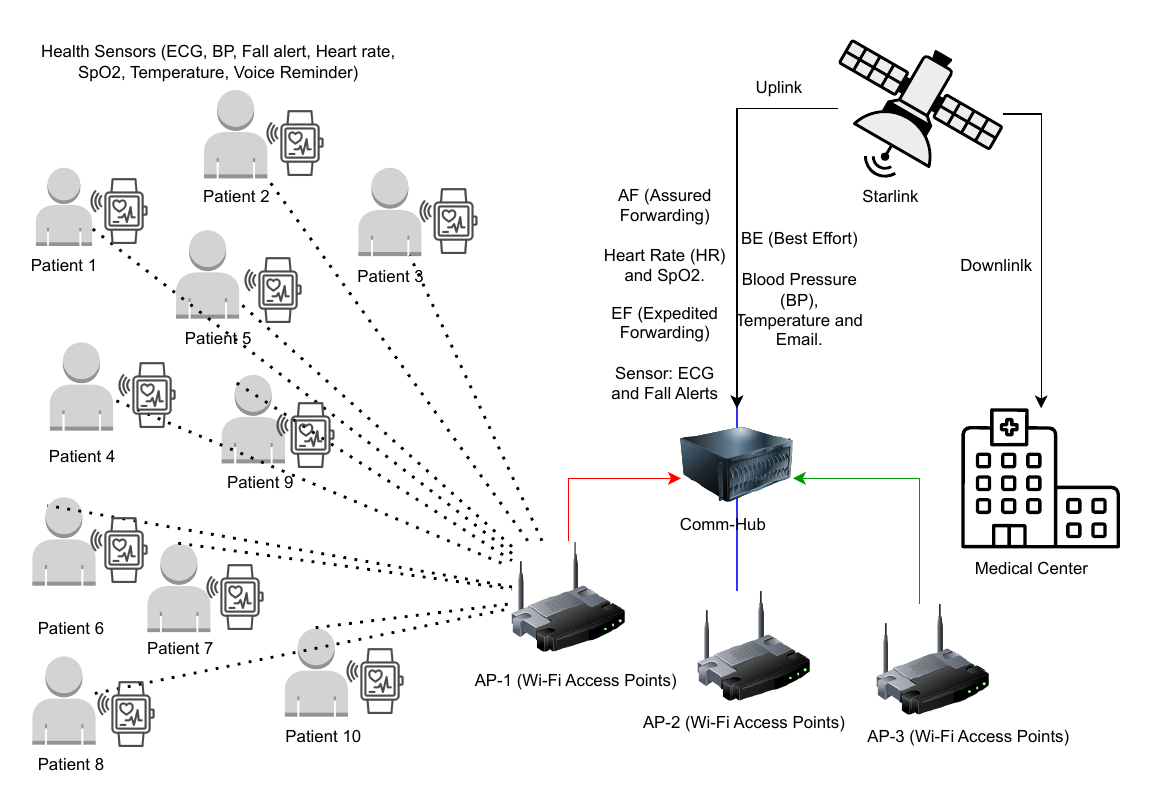}  
\caption{A proposed IoMT-based Healthcare Network}
\label{fig:topology}
\end{figure}
\vspace{-14pt}

\subsection{System Architecture}
The system architecture in Fig. 2 shows the end-to-end workflow of the Starlink-backed healthcare simulation. It includes 15 mobile patient nodes, 3 access points, a hub, a Starlink LEO satellite, a ground station, and a medical center, all using Optimized Link State Routing (OLSR), 802.11n Wi-Fi, and FQ-CoDel-based QoS with DSCP classes (EF: ECG/Fall/Voice, AF: HR/SpO\textsubscript{2}, BE: BP/Temp/Email). Traffic passes through a 50 mbps, 30 ms hub-to-satellite bottleneck and then the starlink path (~41 ms). Voice reminders are sent as EF bursts, and performance metrics (PDR, delay, throughput) are exported to XML for analysis and figure generation.  
\vspace{-10pt}
\begin{figure}[H]
\centering
\includegraphics[height=9.9cm]{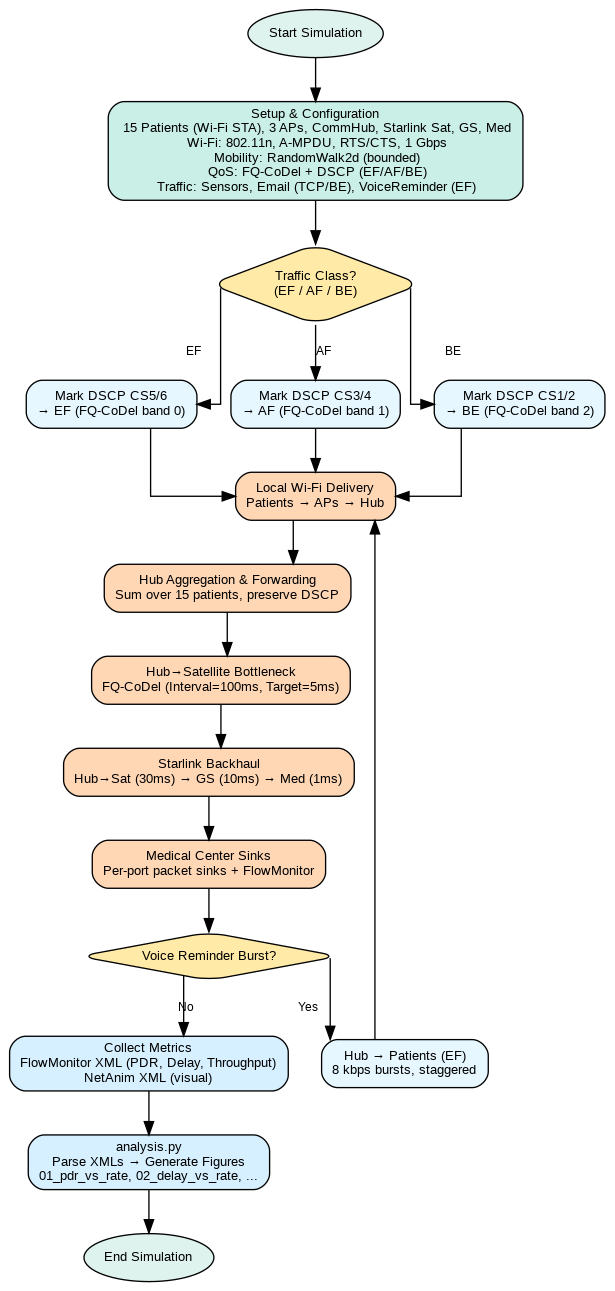}  
\caption{System Architecture Flowchart}
\label{fig:arch}
\end{figure}
\vspace{-10pt}

\subsection{Traffic Types}
The system generates a mix of real-time and non-real-time traffic to reflect the nature of remote patient monitoring. Key sensor data includes ECG, heart rate, SpO\textsubscript{2}, blood pressure, temperature, and fall alerts. ECG and fall alerts are life-critical and time-sensitive, while heart rate and SpO\textsubscript{2} are important but less time-sensitive. Blood pressure and temperature are periodic, non-critical data. The system also handles non-medical traffic like voice reminders and email notifications. Traffic is classified into three DiffServ classes based on urgency.

\begin{table}[H]
\caption{Traffic Classification and DSCP Mapping}
\label{tab:traffic_class}
\centering
\begin{tabular}{|l|l|l|}
\hline
\textbf{Class} & \textbf{DSCP Value} & \textbf{Traffic Types}\\
\hline
EF (Expedited Forwarding) & CS6/CS5 & ECG, Fall/Voice alert\\
AF (Assured Forwarding) & CS4/CS3 & HR, SpO\textsubscript{2}  \\
BE (Best Effort) & CS2/CS1 & BP, Temperature, Email  \\
\hline
\end{tabular}
\end{table}

\subsection{Traffic Control Mechanism}
To prioritize traffic, we use FQ-CoDel at the CommHub to Satellite link. FQ-CoDel offers:
Flow isolation: Prevents bursty flows from starving others.
Bufferbloat control: Maintains low latency under congestion.
Fairness: Prevents any class from monopolizing bandwidth.
Each DSCP class (EF, AF, BE) is assigned a dedicated FQ-CoDel instance to protect EF traffic.

\subsection{Simulation Parameters}
The simulation uses the network simulator (3.45) for modeling Wi-Fi, satellite links, mobility, and QoS mechanisms. Key parameters are as follows:

\begin{table}[H]
\caption{Simulation Parameters}
\label{tab:sim_params}
\centering
\begin{tabular}{|l|l|}
\hline
\textbf{Category} & \textbf{Parameter} \\
\hline
Simulation Duration & 100 s \\
Routing Protocol & OLSR  \\
Patient Nodes  & 15 \\
Wi-Fi Standard & IEEE 802.11n (2.4 GHz) \\
Wi-Fi AP Data Rate & 100 Mbps \\
Wi-Fi AP Delay  & 1 ms \\
Satellite Uplink Bandwidth & 50 Mbps \\
Satellite Downlink Bandwidth  & 1 Gbps \\
ECG Sensor Data Rate  & 16 kbps to 256 kbps \\
Heart Rate Sensor Rate  & 12 kbps \\
SpO\textsubscript{2} Sensor Rate & 8 kbps \\
Blood Pressure Sensor Rate  & 6 kbps \\
Temperature Sensor Rate & 2 kbps \\
Fall Alert Rate & 1 kbps (event-based)\\
Traffic Classes & DSCP(EF, AF, BE) \\
QoS Mechanism & FQ-CoDel (EF/AF/BE) \\
Packet Sizes (ECG/HR/SpO\textsubscript{2}/BP/Temp/Fall):& 80 to 400 bytes \\
\hline
\end{tabular}
\end{table}

\subsection{Simulation Methodology}
The simulation evaluates system performance under varying ECG data rates (16–256 kbps) to assess how the system behaves under increasing load. The methodology begins with the network setup, where a topology is built, IP addresses are assigned, and routing (OLSR) is installed. For mobility, the patient nodes are assigned using the RandomWalk2d model within defined bounds to simulate realistic movement. Applications are configured by using OnOff applications for sensors with DSCP marking, with a VoiceReminder being sent from the CommHub to the patients and an email being sent from the CommHub to the medical center. To ensure quality of service, FQ-CoDel is implemented on the CommHub$\rightarrow$Satellite link. Finally, monitoring is enabled through FlowMonitor and NetAnim to provide detailed analysis of the network’s performance after the simulation.

\subsection{QoS Performance Metrics Evaluation}

\textit{FlowMonitor} sections per-flow statistics (PDR, delay, throughput)
\textit{NetAnim} Visualizes node movement and packet flows
Results exported to \textit{flowmon\_results.csv} for statistical analysis
Graphs comparing PDR, delay, and throughput across different transmission rates are generated for analysis.

\section{Results and Analysis}
This section evaluates the proposed Starlink-Backhaul Smart IoMT Healthcare System for remote patient monitoring in underserved areas. The simulation, conducted using ns-3.45, models a network with 15 mobile patients, three Wi-Fi access points, a communication hub, a Starlink LEO satellite, a ground station, and a medical center, connected via a 50 Mbps backhaul link with 25–35 ms jitter. The system uses FQ-CoDel-based QoS to prioritize critical healthcare data into three classes: EF (real-time, life-critical), AF (important non-real-time), and BE (non-critical). Performance is assessed based on throughput, delay, and packet delivery ratio (PDR) across transmission rates of 16–256 kbps.

\subsection{Throughput Performance}
Fig 3 shows the throughput performance of EF, AF, and BE traffic classes across transmission rates. EF traffic achieves the highest throughput, increasing from 87 kbps at 16 kbps to 500 kbps at 256 kbps, thanks to FQ-CoDel’s prioritization of high-priority flows. AF and BE traffic have lower, nearly flat throughput, peaking at 30 kbps and 50 kbps, respectively, ensuring non-critical data doesn’t consume excessive bandwidth.

\begin{figure}[H]
\centering
\includegraphics[width= 1\linewidth]{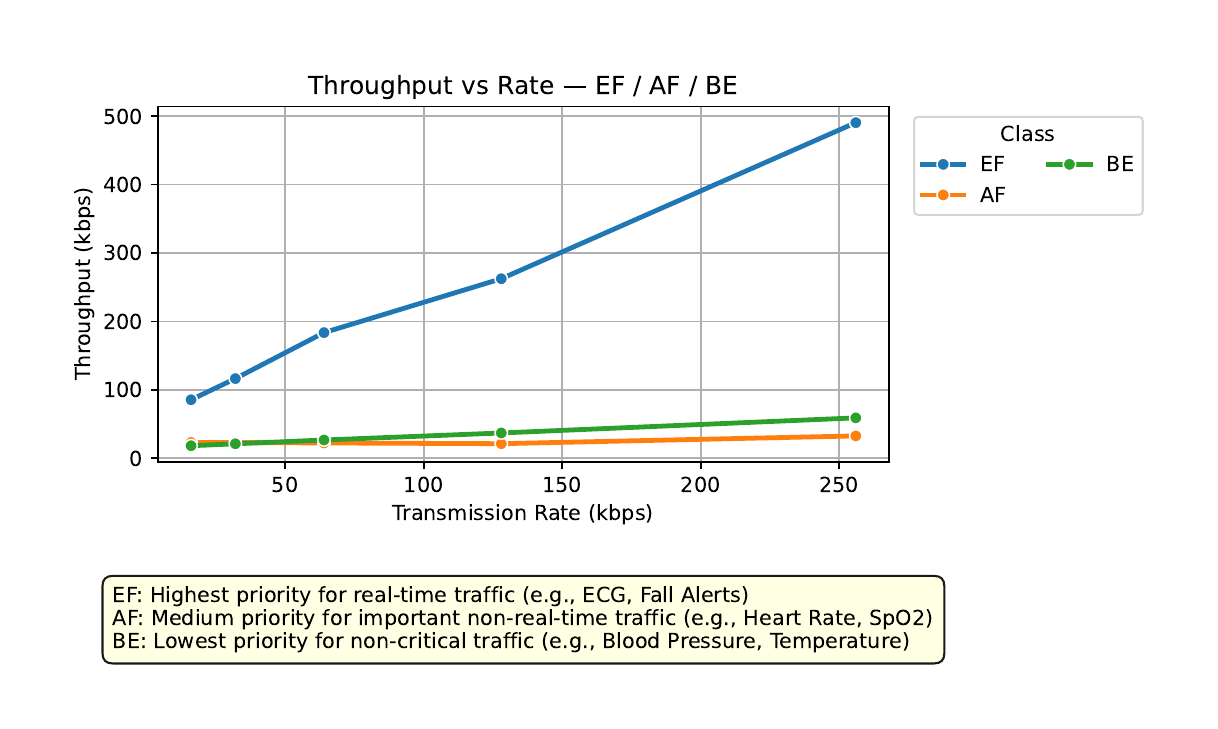}
\caption{Average End-to-End Delay vs Transmission Rate for EF, AF, and BE Classes}
\label{fig:throughput}
\end{figure}

\subsection{End-to-End Delay}
Fig 4 shows that overall delay rises as transmission rates increase because queues build up and links get busier. However, the FDA's latency limit applies only to real-time physiological data in the EF class. In results, EF traffic consistently stays between (5-12 ms), well under the FDA's 125 ms requirement—due to strict prioritization through DSCP tagging and FQ-CoDel management. 
The delay increases seen in AF and BE traffic don't affect FDA compliance, since those classes aren't held to real-time standards. So, even with higher delays in lower-priority traffic, the system still fully meets FDA latency expectations for critical medical data. 
\begin{figure}[H]
\centering
\includegraphics[width= 1\linewidth]{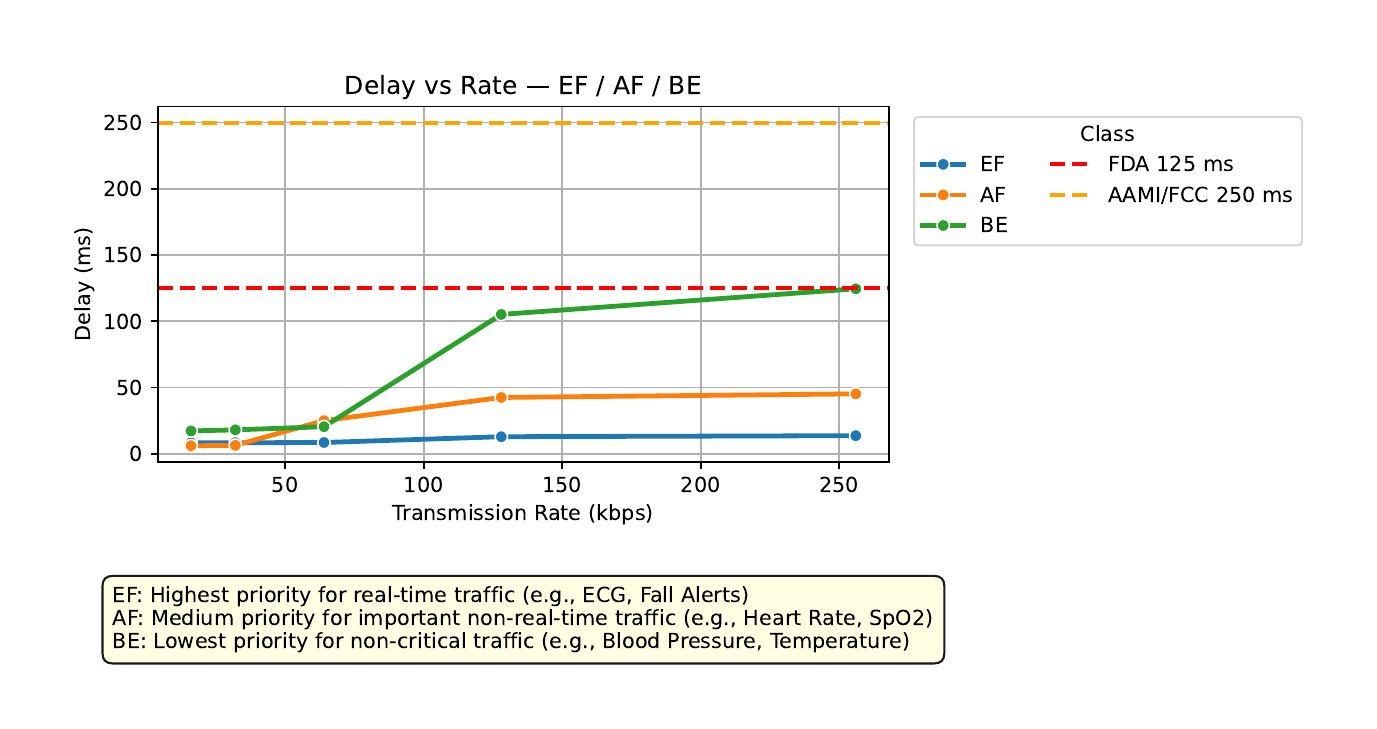}
\caption{Average End-to-End Delay vs Transmission Rate for EF, AF, and BE Classes}
\label{fig:delay}
\end{figure}

\subsection{Packet Delivery Ratio (PDR)}
Fig 5 shows the PDR for each traffic class. At low transmission rates (16–64 kbps), all classes maintain a near-perfect PDR ($>$99\%). As the rate increases, BE traffic drops sharply to 55\% at 256 kbps due to buffer overflow and packet drops, as BE packets are discarded first. EF traffic maintains an 88\% PDR, and AF traffic holds at 67\%, highlighting the QoS mechanism's effectiveness in prioritizing critical data.
\begin{figure}[H]
\centering
\includegraphics[width= 01\linewidth]{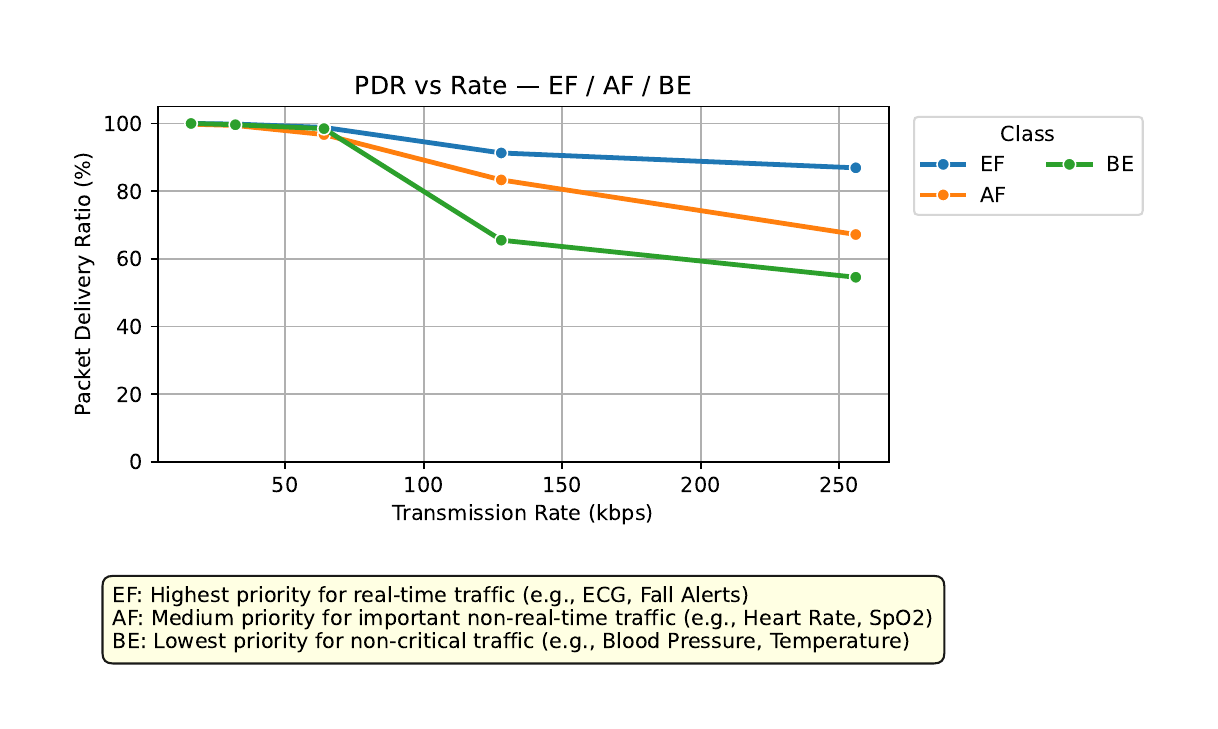}
\caption{Packet Delivery Ratio (PDR) vs Transmission Rate for EF, AF, and BE Classes}
\label{fig:pdr}
\end{figure}

\subsection{Comparison with Prior Work}
\begin{table}[H]
\caption{Performance Comparison with Prior Work}
\label{tab:comparison}
\centering
\begin{tabular}{|l|l|l|l|}
\hline
\textbf{QoS Mechanism} & \textbf{Transmission Model}& \textbf{EF Delay} & \textbf{PDR} \\
\hline
Mamoon et al. \cite{Mamoon2017}& WMMQ&  $>$ 10ms& $\sim$85\%\\
Capozzi et al. \cite{capozzi2023} & PRIO  & 20 ms & 95\% \\
Our Proposed Work& FQ-CoDel  & $\leq$12 ms & $\geq$88\% \\
\hline
\end{tabular}
\end{table}

Our work significantly outperforms prior studies in both latency ($\leq$12 ms vs. $>$95 ms) and QoS enforcement (FQ-CoDel vs. FIFO/PRIO), enabling reliable real-time transmission of critical healthcare data. While the packet delivery ratio (PDR) is lower than that reported by Capozzi et al., this reflects a more realistic and demanding testbed. Our simulation includes 15 patients generating traffic from six sensors each, combined with satellite-induced jitter and bandwidth constraints factors absent in terrestrial-only models. achieving $\leq$12 ms delay and $\geq$88\% PDR for life-critical EF traffic (ECG, fall alerts), outperforming both cognitive radio-based transmission (Mamoon et al., $>$10 ms) and terrestrial LTE (Capozzi et al., 20 ms) models in latency-sensitive scenarios. While BE-class performance degrades under load due to satellite jitter and 15-patient congestion, this trade-off is clinically justified and reflects real-world conditions absent in idealized terrestrial simulations.

This comparison validates Starlink's clinical viability for remote healthcare and sets a new benchmark for QoS-aware satellite telehealth simulations. 
\section{Conclusion and future work }
This study demonstrates that a Starlink-backhauled IoMT healthcare system, enhanced with FQ-CoDel QoS, delivers clinically viable performance for real-time remote monitoring. Our work confirms that Starlink, with edge-based QoS, is not just a connectivity solution it’s a reliable, scalable backbone for equitable, real-time telehealth in underserved regions worldwide. As a natural extension of this work, future research will focus on enhancing resilience and adaptability under dynamic satellite conditions by integrating forward error correction (FEC) to mitigate packet loss without retransmission delay. Dynamic handoff modeling to simulate seamless satellite switching as LEO constellations orbit. Additionally, we plan to incorporate machine learning-based traffic prediction to pre-allocate bandwidth for anticipated critical events (e.g., fall alerts) and evaluate multi-satellite load balancing to distribute patient traffic across beams or satellites, further improving scalability and QoS for large-scale rural deployments.

\section{Acknowledgment}
This research is supported by the Miyan Research Institute, International University of Business Agriculture and Technology, Dhaka 1230, Bangladesh. Research Grant number: IUBAT-MRI-RG-2025.


\begin{thebibliography}{00}

\bibitem{han2022}
Han, L., Retana, A., Westphal, C.,  Li, R. (2023). Large scale leo satellite networks for the future internet: Challenges and solutions to addressing and routing. Computer Networks and Communications, 30-57.


\bibitem{spacex2025}
Sowe, E. A. (2025). Satellite Communications in the New Space Era: A Survey and Future Challenges.

\bibitem{Mamoon2017}
Al Mamoon, I., Muzahidul Islam, A. K. M., Ahmed, A., Baharun, S., \& Komaki, S. (2017). Priority Aware Cognitive Radio Driven Hospital System. \textit{Wireless Personal Communications}, \textit{96}(4), 5973-5994. 

\bibitem{sharma2021}
Sharma, A., Sharma, S.,  Gupta, D. (2021). A review of sensors and their application in internet of things (IOT). International Journal of Computer Applications, 174(24), 27-34.

\bibitem{handley2022}
Izhikevich, L., Enghardt, R., Huang, T. Y.,  Teixeira, R. (2024). A global perspective on the past, present, and future of video streaming over starlink. Proceedings of the ACM on Measurement and Analysis of Computing Systems, 8(3), 1-22.


\bibitem{das2009}
Belgiu, M.,  Drăguţ, L. (2016). Random forest in remote sensing: A review of applications and future directions. ISPRS journal of photogrammetry and remote sensing, 114, 24-31.


\bibitem{sohrabi2020}
Alenoghena, C. O., Ohize, H. O., Adejo, A. O., Onumanyi, A. J., Ohihoin, E. E., Balarabe, A. I., ...  Alenoghena, B. (2023). Telemedicine: a survey of telecommunication technologies, developments, and challenges. Journal of Sensor and Actuator Networks, 12(2), 20.


\bibitem{metz2024}
Hamilton, D., Kohli, S. S. S., McBeth, P., Moore, R., Hamilton, K.,  Kirkpatrick, A. W. (2025). Low Earth Orbit Communication Satellites: A Positively Disruptive Technology That Could Change the Delivery of Health Care in Rural and Northern Canada. Journal of Medical Internet Research, 27, e46113.

\bibitem{Parkhurst2015}
Parkhurst, N. D., Morris, T., Tahy, E., \& Mossberger, K. (2015, May). The digital reality: e-government and access to technology and internet for American Indian and Alaska Native populations. In \textit{Proceedings of the 16th Annual International Conference on Digital Government Research} (pp. 217-229). 

\bibitem{redcross2024}
Kagai, F., Branch, P., But, J., Allen, R.,  Rice, M. (2024). Rapidly deployable satellite-based emergency communications infrastructure. IEEE Access.

\bibitem{alfuqaha2016}
Al-Fuqaha, A., Guizani, M., Mohammadi, M., Aledhari, M., Ayyash, M. (2015). Internet of things: A survey on enabling technologies, protocols, and applications. IEEE communications surveys and tutorials, 17(4), 2347-2376.

\bibitem{John Alexander2022}
Carr, J. A. THE DIGITAL DIVIDE: INTERNET IN ALASKA.

\bibitem{henderson2022}
Sandri, M. (2023). An NS-3 module for non-terrestrial network (NTN) simulation: Implementation, design and performance evaluation. 

\bibitem{capozzi2022}
Sahu, A., Goulart, A.,  Butler-Purry, K. (2016, October). Modeling AMI network for real-time simulation in NS-3. In 2016 Principles, Systems and Applications of IP Telecommunications (IPTComm) (pp. 1-8). IEEE.

\bibitem{ns3consortium2025}
Badini, N., Patrone, F., Zanon, A. M., \& Marchese, M. (2025, March). Evaluating Performance in Satellite Communication Networks: An NS3-Based Simulation Study. In \textit{2025 IEEE Aerospace Conference} (pp. 1-8). IEEE.

\bibitem{capozzi2023}
Michel, F., Trevisan, M., Giordano, D.,  Bonaventure, O. (2022, October). A first look at starlink performance. In Proceedings of the 22nd ACM Internet Measurement Conference (pp. 130-136).




\end{thebibliography}
\end{document}